\documentclass[twocolumn]{aastex631}

\newcommand{\Cx}{{\ensuremath{^{12}\mathrm{C}}}}

\newcommand{\Fe}{{\ensuremath{^{56}\mathrm{Fe}}}}

\newcommand{\He}{{\ensuremath{^{4} \mathrm{He}}}}
\newcommand{\Hy}{{\ensuremath{^{1} \mathrm{H}}}}
\newcommand{\N}{{\ensuremath{^{14} \mathrm{N}}}}
\newcommand{\Ox}{{\ensuremath{^{16}\mathrm{O}}}}

\newcommand{\Si}{{\ensuremath{^{28}\mathrm{Si}}}}
\newcommand{\Ne}{{\ensuremath{^{20}\mathrm{Ne}}}}

\newcommand{\Ca}{{\ensuremath{^{40}\mathrm{Ca}}}}

\newcommand{\Sx}{{\ensuremath{^{32}\mathrm{S}}}}

\newcommand{\Mg}{{\ensuremath{^{24}\mathrm{Mg}}}}

\shorttitle{Magnetic Field in the Primordial Galaxies}
\shortauthors{H.H. Chiu and K.J. Chen}
\graphicspath{{./}{figures/}}
\usepackage{subfigure}

\begin{document}

\title{Magnetic Fields in Primordial Galaxies}

\author{Huai-Hsuan Chiu}
\affil{Department of Physics, National Taiwan University, Taipei 10617, Taiwan}
\affiliation{Institute of Astronomy and Astrophysics, Academia Sinica, Taipei 10617, Taiwan}

\author{Ke-Jung Chen}
\affiliation{Institute of Astronomy and Astrophysics, Academia Sinica, Taipei 10617, Taiwan}



\begin{abstract}

Magnetic fields play a vital role in numerous astrophysical processes such as star formation and the interstellar medium. In particular, their role in the formation and evolution of galaxies is not well understood. This paper presents high-resolution magnetohydrodynamic (MHD) simulations performed with \texttt{GIZMO} to investigate the effect of magnetic fields on primordial galaxy formation. Physical processes such as relevant gas physics (e.g., gas cooling and gas chemistry), star formation, stellar and supernova feedback, and chemical enrichment were considered in the simulations. The simulation results suggest that cosmic magnetic fields can be amplified from $10^{-13}$ G to a few microgauss during cosmic structure evolution and galaxy formation. In the ideal MHD setting, in primordial galaxies at $z>8$, the magnetic energy is less than the thermal and kinetic energy, and therefore, magnetic fields hardly affect the gas dynamics and star formation in these galaxies. Specifically, the consideration of micro-physics properties such as metal diffusion, heat conduction, and viscosity in the MHD simulations, could increase the magnetic field strength. Notably, metal diffusion reduced gas cooling by decreasing the metallicity and thereby suppresses star formation in the primordial galaxies. As a result, the cosmic re-ionization driven by these primordial galaxies may be delayed.
\end{abstract}

\keywords{Cosmology, Galaxy Formation and Evolution, Primordial Galaxies, Cosmic Reionization, Dynamo, Magnetohydrodynamics (MHD)}


\section{Introduction} \label{sec:intro}
Primordial galaxies at redshift $z > 8$ are considered embryos of modern galaxies and the main drivers of cosmic re-ionization \citep{2003.05945,10.1093/mnras/sty1406}, and hence, their formation and evolution is one of the most critical topics in contemporary astrophysics. A promising approach to the study of their formation mechanism is through numerical simulations. Modern hydrodynamics simulations have not only successfully demonstrated the process of individual galaxy formation in a cosmological setting, but also provided insights into the formation of dark matter (DM) halos, galaxies, and large-scale structures. However, information on key physics has been missing in previous studies, and the magnetic field is one of them. Currently, observers can determine the detailed magnetic field structure of galaxies. For example, \cite{fletcher2011magnetic} measured the magnetic field of 21 nearby galaxies and \cite{Haverkorn_2014} suggested that the magnetic field's strength increases toward a galaxy's  center; the magnetic field's strength is independent of the density of the diffuse interstellar gas. \cite{Beck_2015} summarized dozens of magnetic field structures of different galaxy types on the basis of radio observational results.

Star formation, morphology, and outflow of galaxies are influenced by magnetic fields \citep{Shetty_2006,krumholz2019role,Kooij_2021}. Owing to advances in modern supercomputers, cutting-edge magnetohydrodynamic (MHD) simulations have become a powerful tool to explore complex physical processes in galaxy formation. For example, \cite{Su_2017_2} demonstrated that different types of cooling, stellar feedback, and star formation would result in different morphologies of galaxies, and saturation values of magnetic fields. Their simulations of isolated galaxies showed that field strength and gas density were related as $B\propto n^{2/3}$. Furthermore, the magnetic field appears to have only a minor effect in isolated galaxy simulations. \cite{Su_2017} performed high-resolution simulations of isolated galaxies in the Feedback in Realistic Environments (FIRE) Project. They analyzed detailed structures and information about their samples, such as stellar formation history, stellar mass evolution, and morphology, and found that magnetic fields play a minor role compared with turbulence and stellar feedback in galactic star formation. Nevertheless, this does not imply that the role of magnetic fields is negligible in galaxy formation and evolution. Recently, \cite{banfi2021alignment} found that the magnetic field strength determine from Faraday rotation could be underestimated when the magnetic field is aligned with a large-scale filament; their results suggest that magnetic fields may affect the properties of the intergalactic medium.

In this study, we performed high-resolution cosmological simulations to investigate how magnetic fields affect the formation of galaxies and the large-scale structure.  This paper is organized as follows: In Section \ref{sec:method}, we introduce the numerical method used in  our simulations and the halo-finding algorithm used in the present study. In sections \ref{sec:halo} and \ref{sec:gas}, we present the statistical properties of the halo mass function (HMF) and stellar mass function in our simulations. Section \ref{sec:mag} describes our investigation of the amplification of magnetic fields during primordial galaxy formation. Finally, we discuss the implications and applications of our results in Section \ref{sec:discussion} and conclude the paper in Section \ref{sec:result}.

\section{Numerical Method} \label{sec:method}
In this section, we introduce the numerical method used our simulations. We first used the \texttt{MUSIC} code to generate the initial conditions in our cosmological simulations at $z \sim 100$ and then evolved the simulation with the \texttt{GIZMO} code by including the relevant physics required to model galaxy formation (see Section \ref{sec:GIZMO}). Next, following the notation of \cite{Su_2017_2}, we created a suite of models and evolve them up to $z=8$. The names of our models are Hydro, FB, FB+MHD, and FB+MHD+Micro. Their detailed descriptions are presented in Table \ref{tab:label}. Finally, we analyzed DM halos with the Robust Overdensity Calculation using K-Space Topologically Adaptive Refinement (\texttt{ROCKSTAR}) code.  

\begin{table*}[]
\centering
\caption{Names and descriptions of models used in this study.}
\label{tab:label}
\begin{tabular}{ll}
\hline
Model name            & Description                                                                                             \\ \hline
Hydro            & Cosmological hydrodynamics simulation without feedback physics                                                       \\
FB               & Cosmological hydrodynamics simulation with stellar feedback (radiation, stellar winds, supernova feedback)                                        \\
FB$+$MHD         & Cosmological magnetohydrodynamics simulation  with stellar feedback                                                         \\
FB$+$MHD$+$Micro & Cosmological magneto-hydrodynamics simulation  with stellar feedback and diffusion microphysics \\ \hline
\end{tabular}

\end{table*}

\subsection{Cosmic Initial Conditions}
\label{sec:ic}

As mentioned, we generated the initial conditions of our cosmological simulations by using the \texttt{MUSIC} code \citep{Hahn_2011}. The cosmological parameters adopted were $\Omega_0=0.276$, $\Omega_\Lambda=0.724$, $\Omega_b=0.044$, and $h=0.7$, on the basis of WMAP5 data \citep{0803.0547}. We generated a box with a size of 50 Mpc$\, h^{-1}$ at $z = 100$. A higher-resolution region containing a higher density of particles was positioned in the central region with a box size of 10 Mpc$\,h^{-1}$. The finest mass resolution of our simulations could reach $1.82\times 10^5 M_{\odot}$ per gas particle and $9.33 \times 10^6 M_\odot$ per DM particle. Consequently, we had 68.91 million particles representing gas and 68.92 million particles for DM in the higher-resolution region. 
\subsection{\texttt{GIZMO} code} \label{sec:GIZMO}
We used the public version of the GIZMO code along with its meshless finite mass (MFM) algorithm for performing the simulations. The MFM algorithm employs the smoothed-particle hydrodynamics method for covering a large dynamic range and a grid-based scheme for precisely detecting the shocks. The MFM algorithm is an ideal method to model the complex gas dynamics of galaxy formation. The MFM algorithm has been described in detail by \cite{Hopkins_2015_1,Hopkins_2016_1,Hopkins_2016_2}. Numerous test problems have shown that the accuracy and reliability of \texttt{GIZMO} are in good agreement with those of the moving-mesh code \texttt{AREPO} \citep{Springel_2010} and the grid-base code \texttt{ATHENA} \citep{Stone_2008}. In the current study, the prescription of modeling the gas physics followed the implementation of \cite{Hopkins_2014}. Some of the key features of our simulations are summarized in the following paragraphs.

\subsubsection{Gas Cooling and Star Formation}
\label{sec:2-1}
The cooling module in \texttt{GIZMO} performs the heating and cooling of H and He through photonionization, recombination, collision, and free-free emission. The cooling curves were adopted from \cite{Wiersma_2009} and they corresponded to cooling rates from $10$ K to $10^{10}$ K, involving 11 chemical species \footnote{Isotopes: \Hy, \He, \Cx, \N, \Ox, \Ne, \Mg, \Si, \Sx, \Ca, \Fe}. The redshift-dependent UV background used for heating was based on  \cite{Faucher_Gigu_re_2009}. 
In the simulations, star formation could occur only in a  dense, molecular, self-gravitating region where the gas density exceeded the critical density $n_{crit}$ of 100 cm$^{-3}$ and the self-gravitating criterion presented by \cite{Hopkins_2013} was fulfilled. Furthermore, star formation occurred only in the molecular gas, whose $f_{H_2}$ fraction was determined by the local density and metallicity and which was formulated by \cite{Krumholz_2011}.  If all of the above criteria are met, stars are formed at a rate $\dot{\rho}_\star\propto\rho_{mol}/t_{f}$, where $\rho_{mol}$ is the molecular density and $t_f$ is the free-fall time. Star formation simulations of \cite{Hopkins_2011,Hopkins_2012,Hopkins_2013} showed that as long as stellar feedback is provided and the largest fragmentation scales in galaxies are well resolved, this star formation module can reproduce the observed galactic star formation rate and history.

\subsubsection{Stellar Feedback}
\label{sec:feedback}
Once a star particle forms, it is treated as a single stellar population with known age, mass, and metallicity. The stellar feedback of star particles is given by \texttt{STARBURST99} \citep{Leitherer_1999}, with the standard initial mass function being assumed \citep{Kroupa_2002}. The relevant feedback processes considered in the simulations were as follows:
\begin{enumerate}
\item Radiation Pressure: Stellar radiation expels gas from the star formation region and injects momentum to the interstellar medium. We considered local and long-range momentum deposition by radiation pressure by considering the initial UV and optical flux as well as multiple scattering of re-emitted IR flux.

\item Photo-ionization and Photo-electric Heating: Once stars form, they emit copious UV photons that ionize and heat their surrounding gas. This process also alters the heating and cooling of the gas.

\item Stellar Winds :  Evolved stars such as O-type and asymptotic giant branch stars can have strong stellar wind, which posit a strong stellar feedback. Our feedback model included the mechanical power of stellar wind tabulated as a function of the star's age and metallicity \citep{10.1111/j.1365-2966.2012.20593.x}. 

\item Supernovae of Type Ia and II : When stars die as supernovae, they deposit ejecta energy, momentum, mass, and metals in the surrounding gas. The mechanical and chemical feedback of supernovae have been presented by \cite{Leitherer_1999} and \cite{Mannucci_2006}.   
\end{enumerate}

\subsubsection{Magnetic Fields}

The MHD module coupled with the MFM algorithm in \texttt{GIZMO} can capture complex phenomena \citep{Hopkins_2015_2} such as magnetorotational instability, the launching of magnetic jets, and mixing driven by both Rayleigh–Taylor and Kelvin–Helmholtz instabilities. Furthermore, to eliminate the non zero $\nabla\cdot B$, the \texttt{GIZMO} code uses a combination of the cleaning methods of \cite{POWELL1999284} and \cite{DEDNER2002645}. In our MHD models, we added an initial uniform magnetic field of $1.73 \times 10^{-14}$ G at $z=26$ since previous studies \citep{Marinacci2015} have shown that the final strength of galactic magnetic fields is not sensitive to the value of seed fields as long as the comoving magnetic field is below the value of $10^{-9}$ G when the initial magnetic field is seeded. In our simulation, we considered the magnetic field at the ideal-MHD limit.

\subsubsection{Magnetic-field-related Micro-physics}
\label{sec:microphysics}
Several micro-physical processes were included in \texttt{GIZMO}  and their implementations have been descibed by \cite{Su_2017}. Here, we summarize the key features of the microphysics considered in our runs.\\
\newline
\textbf{Anisotropic Conduction} 
To consider thermal conduction, we added an extra diffusion term in the ideal MHD equations \citep{PhysRev.89.977}. We calculated the conduction coefficient self-consistently as the Spitzer conductivity, considering the ionized fraction, electron mean free path and the temperature gradient scale length. In the expression, the thermal conduction was anisotropic but aligned with the magnetic field lines \citep{Hopkins_2016_3}. Hence, the evolution of the magnetic field was critical to the conduction effect.\\ 
\newline
\textbf{Anisotropic Viscosity}
Viscosity was incorporated into the MHD equations through the Navier-Stokes equations by modifying the momentum and the energy flux in the Euler equations. The viscous coefficients were calculated self-consistently as the Braginskii viscosity \citep{1965RvPP....1..205B}, considering the average ion mass, ion mean free path, and scale length of velocity. Similar to the case of conduction, the presence of a magnetic field destroys the isotropy of viscosity.\\
\newline
\textbf{Metal Diffusion}
On the basis of the studies of  \cite{Colbrook_2017}, \cite{Hopkins_2018}, and \cite{Rennehan_2018}, the diffusion scheme used in the current study assumed that for a scalar $A$, the diffusion time derivative was $\nabla\cdot (K\nabla A)$, where the diffusion coefficient $K$ is given by $C|S|h^2$; here, $|S|$ is the norm of the trace-free shear tensor, $C$ is a dimensionless constant set in the parameter file, and $h$ is the resolution scale.

\subsection{Dark Matter Halo Finding}
We use \texttt{ROCKSTAR} \citep{Behroozi_2012} for data analysis and to search for galactic halos in our simulations. The halo-finding method was based on the adaptive hierarchical refinement of friends-of-friends (FOF) groups, and it involved the consideration of six phase-space dimensions (three spatial dimensions and three momentum dimensions) and one time dimension. \texttt{ROCKSTAR} uses the phase-space information accurately to define which halo every single DM particle should belong to, and it defines the host halo as the most massive halo in every FOF group. This definition is similar to that of \cite{Pillepich2017a}, which also used the halo mass to define the host-sub halo relationship. 
Finally, we reassigned star particles back to each (sub)halo on the basis of the viral radius of each halo. 


\section{Statistical Properties of DM Halos and Galaxies } \label{sec:halo}
In this section, we present the HMF and the stellar mass function obtained from our simulations and compare them with those of previous studies. Such a comparison would serve as a consistency check of our simulations in terms of the halo statistics and properties.
\begin{table}
\caption{Resolutions of GIZMO and TNG50-2}
\label{tab:resolution}
\begin{tabular}{|l|l|l|}
\hline
                  & TNG50-2  & \texttt{GIZMO}   \\ \hline
Boxsize (comoving) & 35 $Mpc\,h^{-1}$ & 50 $Mpc\,h^{-1}$ \\ \hline
$M_{gas}$            & $6.8\times 10^5$    & $2.6\times 10^5$  \\ \hline
$M_{dm}$             & $3.6\times 10^6$    & $1.3\times 10^6$  \\ \hline
\end{tabular}
\end{table}
\begin{figure}
    \centering
    \includegraphics[width=0.9\linewidth]{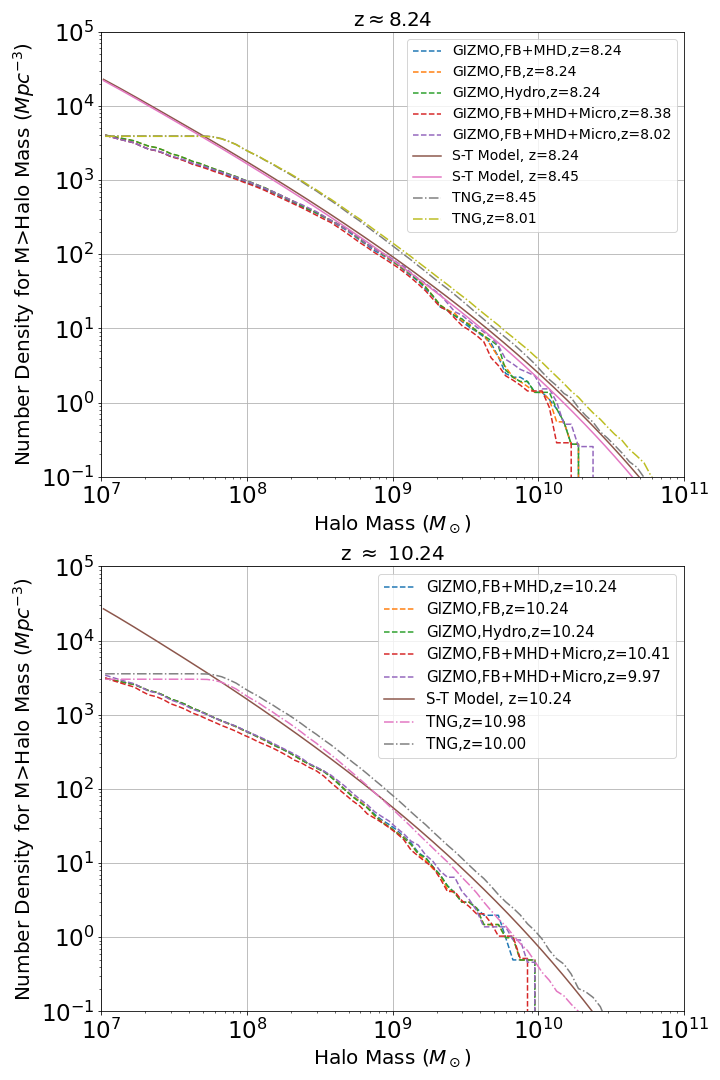}
    \caption{HMF at $z=10.24$ (lower panel) and $z=8.24$ (upper panel). For halos with $10^8 - 10^9 M_\odot$, the HMF of our models fitted well with the IllustrisTNG simulation and the S-T model. The deviation at the low-mass end ($<10^8M_\odot$) was caused by the simulation resolution, while at the high-mass end ($>10^8M_\odot$), it was associated with the size of the simulation box.}
    \label{fig:HMF}
\end{figure}

\begin{figure}
    \centering
    \includegraphics[width=\linewidth]{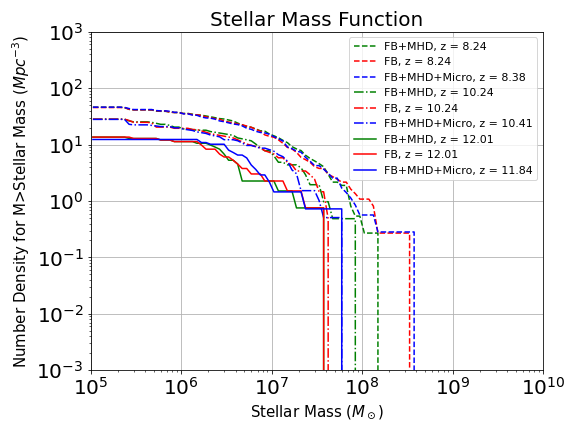}
    \caption{Stellar mass function. Different colors are used for different models(red for FB, green for FB+MHD, blue for FB+MHD+Micro), and different line styles represent different redshifts. The high-mass ends of the FB+MHD+Micro model at  $z=10.41$ and $z=11.84$ are indistinguishable, and the lines of $z=12.0$ for the FB+MHD and FB models overlap. For a given redshift, the profiles of the three models are similar.}
    \label{fig:SMF}
\end{figure}
\begin{figure}[h]
    \centering
    \includegraphics[width=\linewidth]{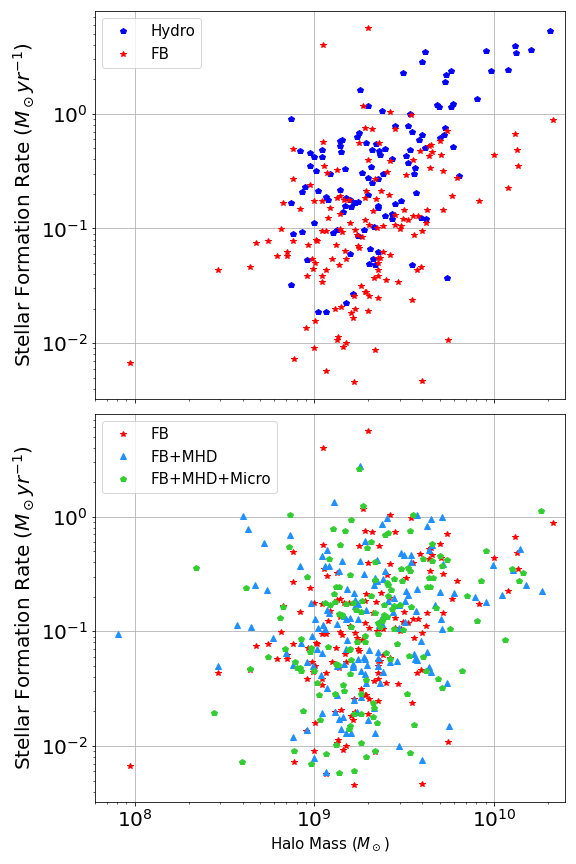}
    \caption{SFR at $z=8.3$. Upper panel: Comparison between the Hydro and FB models. Stellar feedback suppressed star formation for halos with a mass exceeding $10^{9} M_\odot$. Lower panel: Comparison between the FB, FB+MHD, and FB+MHD+Micro models. The scattered data points are located roughly in the same region, and it is difficult to find differences between the models. }
    \label{fig:SFR}
\end{figure}
\subsection{Halo Mass Function} \label{sec:HMF}
Figure \ref{fig:HMF} presents the HMF obtained from our simulations along with public data from the IllustrisTNG Project \citep{Vogelsberger2014a,Vogelsberger2014b}. The IllustrisTNG Project is series of MHD cosmological simulations that models galaxy formation and evolution with the classic $\Lambda$CDM cosmological model. The Illustris TNG Project provided a benchmark models for our simulations. 
As shown in Figure \ref{fig:HMF}, our results matched well with the TNG50-2, whose resolution and box size were similar to those used in our simulations. However, the IllustrisTNG Project\citep{Weinberger2016,Pillepich2017b} considered black hole feedback and galactic winds, which were neglected in our runs.
\par Our \texttt{GIZMO} runs had a smaller number density of halos than IllustrisTNG at high redshift, as shown in the bottom panel of Figure \ref{fig:HMF}. Halos of $10^9 M_\odot$  ($2\times 10^8$ to $1.5\times 10^9 M_\odot$ for $z\approx8$) were consistent with both theoretical prediction and IllustrisTNG simulation. The gravitational potential wells of these halos were sufficiently deep to prevent the expulsion of the majority of the gas from the host halo. We also compared the HMF of our simulations with the Sheth-Tormen model \citep{Sheth2002}\footnote{It is to be noted that since observations cannot probe the HMF at such a high redshift, both the analytical model and IllustrisTNG results served as our benchmarks.}. With the \texttt{GIZMO} data sets, the HMF was similar among the four models, showing that the stellar feedback, magnetic field, and other micro physics described in Section \ref{sec:microphysics} slightly affected the large-scale structure. Our results differed from both the S-T model and IllustrisTNG data at both high-mass and low-mass ends. At the low-mass end, given the mass resolution in Table \ref{tab:resolution}, every DM halo can be resolved by at most 100 DM particles. On the other hand, at the high-mass end, the high-resolution region of our run was 10 Mpc, which limited the formation of massive halos. Consequently, the halos around $10^8 - 10^9 M_\odot$ were fairly consistent with the S-T model and IllustrisTNG data.

\begin{figure*}
    \centering
    \includegraphics[width=\linewidth]{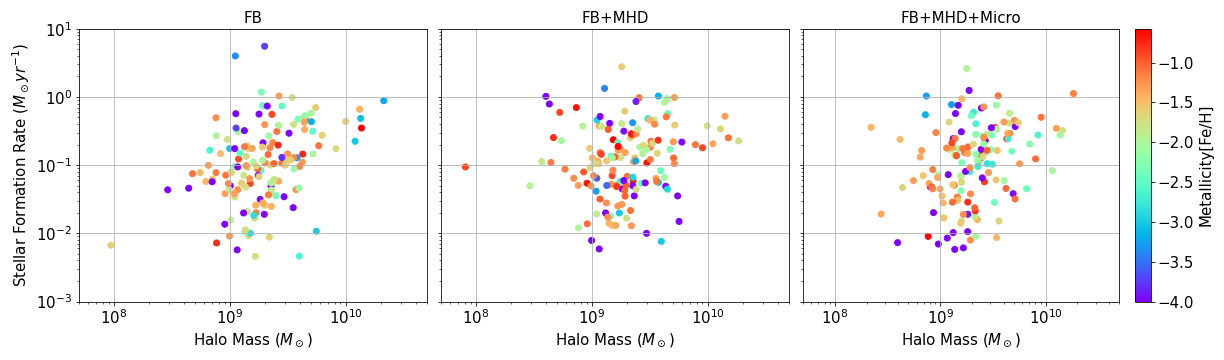}
    \caption{Correlation between the SFR and gas metallicity. The color of the data points indicates the [Fe/H] value, which represents the gas metallicity. The low-metallicity population is distributed around the halo mass of about $10^9 M_\odot$. Data in the central panel, which represents the FB+MHD model, are redder than those of the other models, implying that the presence of a magnetic field can prevent metal diffusion. (see Section \ref{sec:microphysics}).}
    \label{fig:SFR_Metal}
\end{figure*}
\begin{figure}[h]
    \centering
\includegraphics[width=0.8\linewidth]{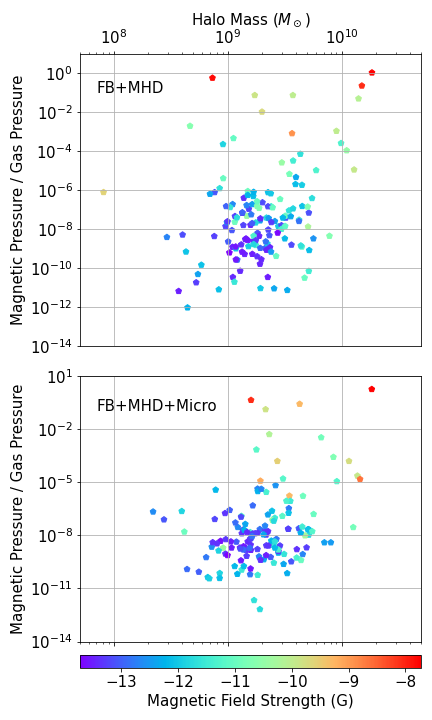}
    \caption{Ratio between the gas thermal energy density and magnetic energy density. The color of data points represents the magnitude of the magnetic field. The magnetic energy density in most of our simulated galaxies was less than 1\% of the gas internal energy. A few galaxies that had a magnetic energy comparable to the thermal energy are discussed in Section \ref{sec:morphology}.}
    \label{fig:MagEnergy}
\end{figure}
\subsection{Stellar Mass Function}
To better quantify the halo properties, we plot the corresponding stellar mass function of luminous halos in Figure \ref{fig:SMF}. For profiles around $10^7M_\odot$, three models (FB, FB+MHD, and FB+MHD+Micro) showed differences at high redshift, while they converged to the same profile around $z=8.3$. It is noteworthy that the volume-weighted average magnetic field strength around $z=8.3$ was only around $10^{-10}$ G. According to \cite{Marinacci2015}, the magnetic field hardly affects the gas dynamics until the magnetic field pressure becomes comparable to the thermal gas pressure. For this criterion to be fulfilled, the magnitude of the magnetic field strength should exceed $10^{-9}$ G. Hence, the similarity among models is not surprising since the magnetic field played only a minor role in our MHD simulations (further discussion is presented in Section \ref{sec:magenergy}).

\section{Gas physics and star formation}
\label{sec:gas}

As discussed in Section \ref{sec:intro}, the magnetic pressure may provide additional support against the gravitational collapse of gas and reduce the star formation rate (SFR). In this section, we further quantify the influence of the magnetic field on the gas physics and star formation in primordial galaxies. In the following section, we compare only the FB, FB+MHD and FB+MHD+Micro model, focusing on the magnetic field's effect and the microphysics.

\subsection{Star Formation Rate and Metallicity}
We present our simulation around $z=8.3$. In Figure \ref{fig:SFR},  most of the halos are located at the mass of $\approx 10^9 M_\odot$, which is consistent with the analysis of HMF. The SFR dependence of the halo mass is unclear from our simulations owing to the data points being scattered. In the upper panel, it is clear that stellar feedback can suppress the overall SFR in galaxies with masses below $10^9 M_\odot$ . Furthermore, the difference among the FB, FB+MHD, and FB+MHD+Micro models is small as shown in the lower panel of Figure \ref{fig:SFR}.

In Figure \ref{fig:SFR_Metal}, we plot the three models separately; the color indicates the gas metallicity of halos. The purple points are halos with low metallicity, and they can be considered as belonging to primordial galaxies. This indicates that primordial galaxies were mainly formed around the halo mass of $10^9 M_\odot$ in our simulations. There were 18 galaxies in the FB+MHD data with [Fe/H]$>0.1$ \footnote{We define the metallicity as [Fe/H], which is the common logarithm of the mass ratio of Fe to H of the gas compared to that of the Sun.}, while there were only 10 (FB) and 12 (FB+MHD+Micro) for other two models. In other words, the galaxies of the FB+MHD model showed higher metallicity than those of the FB and FB+MHD+Micro models, implying that the magnetic field may confine heavy elements and prevent them from being expelled from the host halos after supernovae feedback. 
\begin{figure*}
    \centering
    \includegraphics[width=0.75\linewidth]{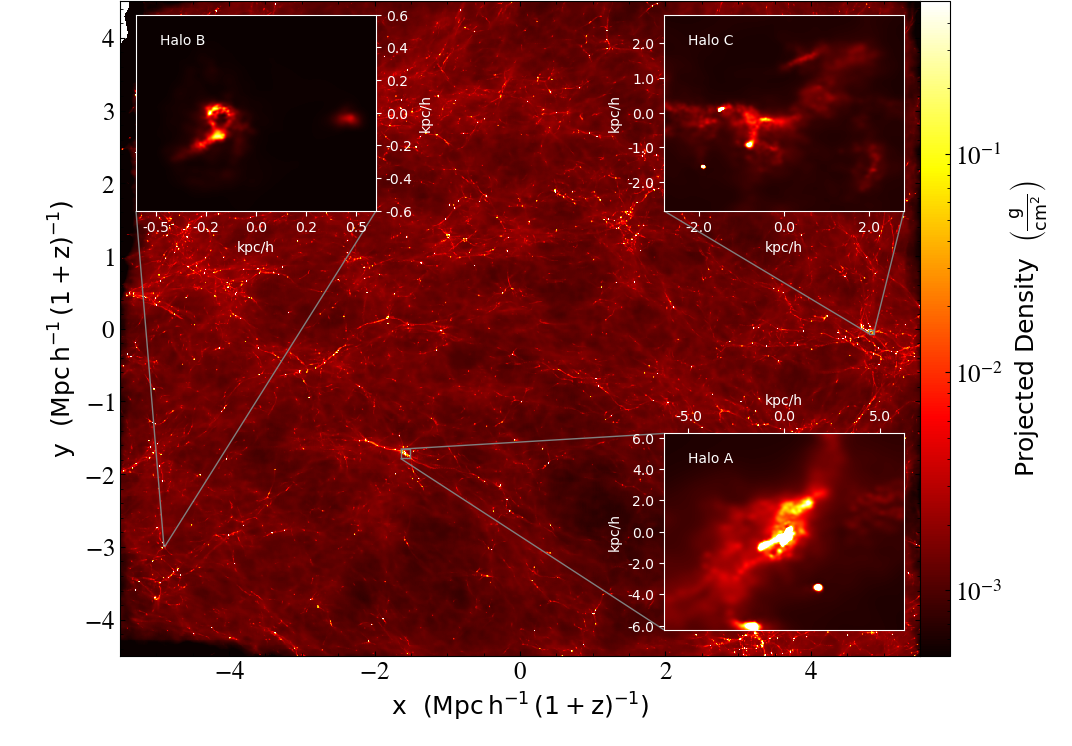}
    \caption{Projected density plot of the FB+MHD model at $z=8.25$. The figure also shows magnified images of three halos in which the magnetic energy density exceeded 10\% of the gas thermal energy.}
    \label{fig:fancy_21}
\end{figure*}
\begin{figure*}
    \centering
    \includegraphics[width=0.75\linewidth]{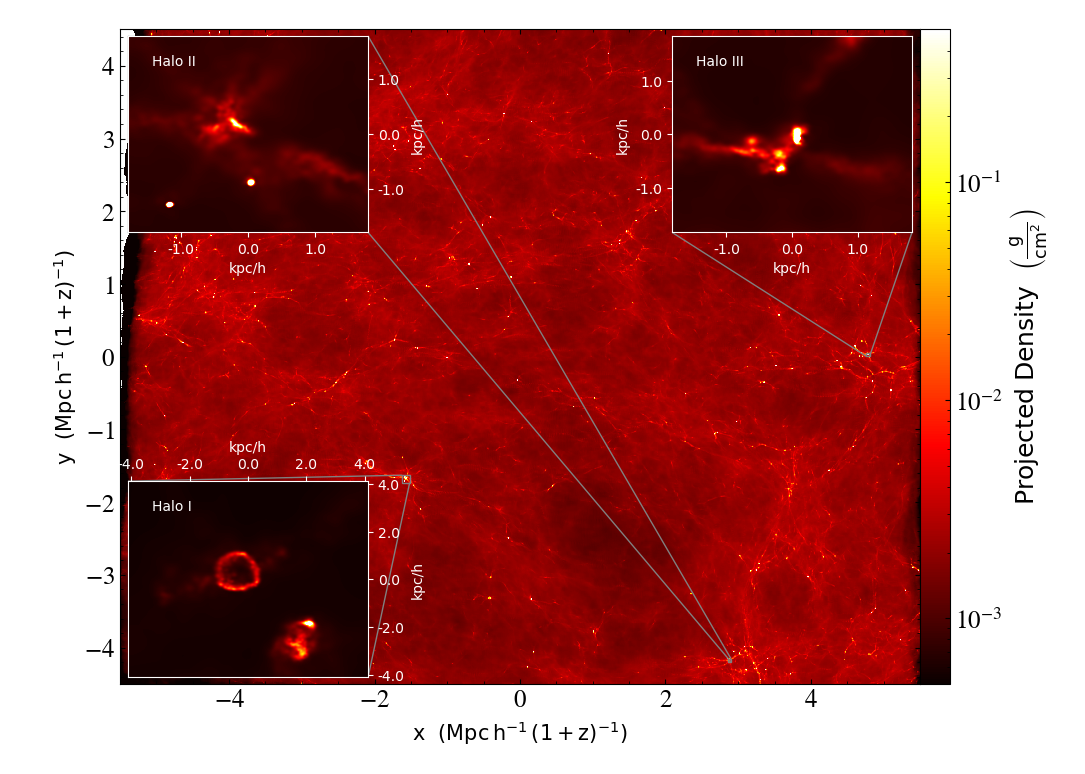}
    \caption{Projected density plot of the FB+MHD model at $z=10.24$. Halos I, II, and III were the three heaviest halos in our simulations. Halos I and III were the progenitor halos of Halos A and C shown in Figure \ref{fig:fancy_21} because of their position. A comparison with the magnified image in Figure \ref{fig:fancy_21}, suggests that Halo I may have just undergone a supernova explosion at this $z$.}
    \label{fig:fancy_17}
\end{figure*}

\subsection{Magnetic Energy} \label{sec:magenergy}
In the previous section, we showed that differences in the metallicity and SFR among our models were mainly associated with stellar feedback. Adding a magnetic field or micro physics affected the results only slightly, mainly because most of the halos in our simulations had a weak magnetic energy, which was not comparable with the internal energy. In Figure \ref{fig:MagEnergy}, we plot the ratio of the internal energy density to the magnetic energy density as a function of the halo mass. In our simulations, the magnetic energy of most of the galaxies was below $1 \%$ of their internal energy. The magnetic energy density in a few galaxies could become comparable with the internal energy density because of the dynamo effect. Figure \ref{fig:MagEnergy} explains why the metallicity and SFR distribution were almost identical among different models in previous sections: the magnetic field was too weak to compete with the thermal energy and therefore did not significantly influence the gas dynamics. Our simulation results showed that the gas thermal energy and kinetic energy dominated over the magnetic energy at star formation sites of primordial galaxies.

\section{Evolution of Magnetic Field}\label{sec:mag}
Previous simulations \citep{2010.09729} have shown that the spatial resolution in MHD simulations are critical for the amplification of magnetic fields since the turbulent dynamo grows considerably faster on smaller scales. Hence, even state-of-art zoom-in simulations underestimate the amplification of the magnetic field. Even though, the comparison of our simulations with and without microphysics remains robust.

\subsection{Morphology} \label{sec:morphology}
Figure \ref{fig:fancy_21} shows the density projection plot of our FB+MHD model at $z=8.25$. Three halos are evident: Halos A, B, and C; their magnetic energy density is comparable to their thermal energy density. Halo A is also the most massive halo in this snapshot. These magnified figures of halos show that their morphology is irregular. None of them is spherically symmetric. A plot similar to that of Figure \ref{fig:fancy_21} but for $z=10.24$ is shown in Figure \ref{fig:fancy_17}. At $z=10.24$ , the magnetic energy density in all halos was considerably less than the thermal energy density. Hence, we identified the three most massive halos at this redshift. On the basis of their locations, the Halos I and III correspond to Halos A and C in Figure \ref{fig:fancy_21}, respectively. The ring structure of Halo A suggests that it could have undergone supernova explosions. Furthermore, the structure of Halo C is largely unchanged during its evolution from $z=10.24$ to $z=8.25$. The morphology of the halos suggests that the spherical distribution of the halo gas broke down in these primordial galaxies.

\begin{figure}
    \centering
    \includegraphics[width=\linewidth]{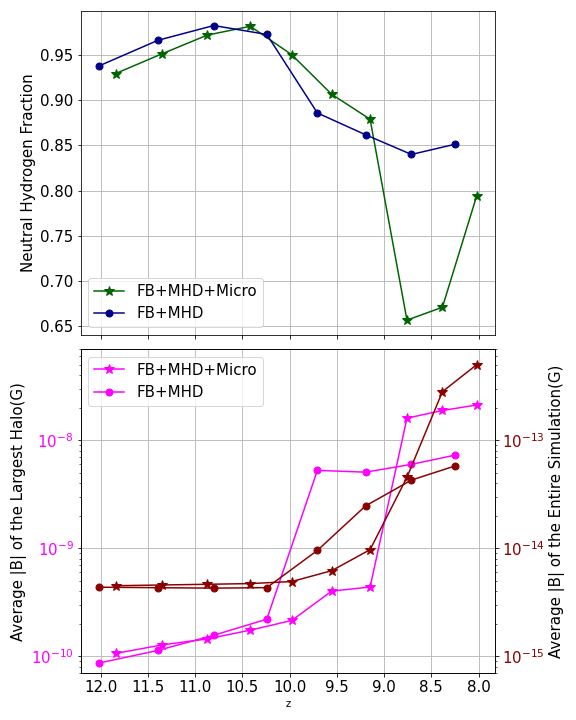}
    \caption{Upper panel: Average neutral hydrogen fraction in the heaviest halo as a function of the redshift. Lower panel: Evolution of the magnetic field magnitude of the most massive halo and the cosmic volume. The brown curve shows the average magnetic field magnitude as a function of the red-shift, and it corresponds to the y-axis on right-hand side. The pink curve indicates the evolution of the average magnetic field magnitude of the most massive halo and the values are presented on the left y-axis. Clearly, when the gas is violently ionized(for example, $z$ from  10.2 to 9.8 in the FB+MHD case, shown by the red line in the upper panel), the corresponding magnetic field strength increases significantly.}
    \label{fig:bfield}
\end{figure}
\subsection{Magnetic Field Strength} \label{sec:mag_strength}
In this section, we calculate the volume-weighted average magnetic field of the central domain with a size of 10 comoving Mpc$\,h^{-1}$ , containing the high-resolution region considered in our simulations in Figure \ref{fig:bfield}.
\par First, the growth magnitude of the magnetic field in the halo was slightly larger than that of the cosmic volume. This is understandable since the magnetic field is mainly amplified inside galaxies. Second, the magnetic field amplification for the FB+MHD+Micro model occurred later than that for the FB+MHD model. This result is consistent with the evolution of the neutral gas fraction in the heaviest halo, since metal diffusion suppresses the gas cooling of galaxies and the consequent star formation, thereby finally delaying the low neutral gas fraction to a later time. Since magnetic field amplification is driven by the turbulence of ionized gas, the fraction of ionized gas correlates well with the amplification of the magnetic field. 
\par On the other hand, although the magnetic field of the FB+MHD+Micro model increased later, its final magnitude was larger than that of the FB+MHD model at a lower red-shift. This can be explained by the effects of anisotropic conduction and viscosity. These micro-physics processes enhance gas motion, which leads to greater amplification of the field strength by a factor of two or three.
\par Last but not least, Figure \ref{fig:bfield} shows the magnetic field strength as a function of the halo mass.  Starting with $z \sim 12$, the field strength first increased with the red-shift slowly, but showed a sudden jump by a factor of 100 around $z \sim 9 - 10$. Compared with the evolution of ionized gas fraction (see upper panel of Figure \ref{fig:bfield}), the sudden jump in the magnetic field occurred when the fraction of ionized gas increased. This indicates that the magnetic field amplification process in primordial galaxies is associated with the ionization state of the gas.

\section{Discussion} \label{sec:discussion}
We discuss the physics underlying our simulation results and their implications regarding cosmology. In Section \ref{sec:gas}, we analyzed the relation between the SFR, halo mass, and metallicity. The magnetic field played only a minor role in our simulations, mainly because the thermal and kinetic energies of gas dominated over the magnetic field. Later, we analyzed the evolution of the magnetic field for the largest halo and found that the micro physics might have affected the ionization and amplification of the magnetic field. (see Section \ref{sec:mag_strength})
\par Our simulations differ from those of previous works, such as IllustrisTNG and the FIRE Projects \citep{Su_2017}. In the IllustrisTNG Project, a series of MHD simulation with different resolutions were conducted without including the microphysics mentioned in Section \ref{sec:microphysics}. On the other hand, \cite{Su_2017} performed simulations in which microphysics was considered. However, they focused on zoom-in simulations of isolated galaxies and targeted low-redshift galaxies. Through cosmological simulations, our research could directly trace primordial galaxy formation by modeling the co-evolution of a large-scale structure and the associated galaxy formation. 
\par Although we have attempted to push the envelope of cosmological MHD simulations, our simulations still suffer from a few technical limitations. In previous sections, we have suggested that the stellar formation rate and halo mass appears to be uncorrelated at redshift $z \ge 8$. This may be a bias associated with the halo mass of our samples; our dataset mainly comprised sample halos around $10^9 M_\odot$, and it is limited by the boxsize and mass resolution. Another issue is the strength of the magnetic field. The magnetic field starts to affect the halo gas only when the magnetic pressure is comparable to the thermal pressure. However, since the amplification of the magnetic field is sensitive to the simulation resolution, higher resolution simulations are required to pinpoint the role of the magnetic field in primordial galaxies.

\section{Conclusion} \label{sec:result}
We performed cosmological MHD simulations using the \texttt{GIZMO} code, with the mass resolution reaching $10^5-10^6 M_\odot$ per gas and DM particle and with a kilo-parsec-scale spatial resolution. Our suite of simulations involved four models, to examine the effect of stellar feedback, magnetic field, and other magnetic-field-related micro-physics such as conduction and viscosity. We successfully reproduced the halo mass function, which was consistent with previous work. Furthermore, stellar feedback dominated the SFR in galaxies, and our result agrees with those of previous simulations \citep{Su_2017}, validating their argument of stellar feedback at both low and high redshifts. However, if we include the micro-physics such as metal diffusion, heat conduction, and viscosity in the MHD simulations, they can enhance the magnetic field strength. More importantly, the microphysics reduces gas cooling by decreasing the metallicity and hinders star formation in primordial galaxies. The consequent cosmic reionization driven by these primordial galaxies may be delayed to a lower redshift.


\begin{acknowledgments}
This work is  supported by the Ministry of Science and Technology, Taiwan under grant no. MOST 110-2112-M-001-068-MY3 and the Academia Sinica, Taiwan under a career development award under grant no. AS-CDA-111-M04. HC thanks Sean Chao and Yi-Kuan Chiang for their technical support.
KC thanks the hospitality of the Aspen Center for Physics, which is supported by NSF PHY-1066293, and the Kavli Institute for Theoretical Physics, which is supported by NSF PHY-1748958, the Nordic Institute for Theoretical Physics (NORDITA), and Physics Division at the National Center for Theoretical Sciences (NCTS). Our computational resources are provided by the National Energy Research Scientific Computing Center (NERSC), a U.S. Department of Energy Office of Science User Facility operated under Contract No. DE-AC02-05CH11231, and the TIARA Cluster at the Academia Sinica Institute of Astronomy and Astrophysics (ASIAA).
\end{acknowledgments}

\end{document}